\documentclass[sigconf]{acmart}
\AtBeginDocument{%
  }

\setcopyright{acmlicensed}
\copyrightyear{2018}
\acmYear{2018}
\acmDOI{XXXXXXX.XXXXXXX}
\acmConference[Conference acronym 'XX]{Make sure to enter the correct
  conference title from your rights confirmation email}{June 03--05,
  2018}{Woodstock, NY}
\acmISBN{978-1-4503-XXXX-X/2018/06}

\begin{document}

\title[AI-Mediated Serious Illness Conversations in the ED]{Exploring the Feasibility and Acceptability of AI-Mediated Serious Illness Conversations in the Emergency Department}

\author{Hasibur Rahman}
\authornote{Both authors contributed equally to this work}
\orcid{0009-0008-0938-1504}
\affiliation{
  \institution{Northeastern University}
  \city{Boston}
  \state{Massachusetts}
  \country{USA}
}
\email{rahman.has@northeastern.edu}

\author{Kenji Numata}
\authornotemark[1]
\orcid{0000-0001-9813-5592}
\affiliation{
  \institution{Sei Marianna Ika Daigaku}
  \city{Kawasaki}
  \country{Japan}
}

\author{Evelyn T. Lai}
\orcid{0009-0002-8897-0441}
\affiliation{
  \institution{Department of Emergency Medicine, Brigham and Women’s Hospital}
  \city{Boston}
  \state{Massachusetts}
  \country{USA}
}
\email{etlai@bwh.harvard.edu}

\author{Maria Cheriyan}
\orcid{0009-0002-8747-7663}
\affiliation{
  \institution{Harvard College}
  \city{Boston}
  \state{Massachusetts}
  \country{USA}
}
\email{mcheriyan@college.harvard.edu}

\author{Adrian Haimovich}
\orcid{0000-0002-4106-7055}
\affiliation{
  \institution{Beth Israel Deaconess Medical Center}
  \city{Boston}
  \state{Massachusetts}
  \country{USA}
}

\author{Kei Ouchi}
\authornote{Both authors contributed equally to this work}
\orcid{0000-0002-2723-4259}
\affiliation{
  \institution{Department of Emergency Medicine, Harvard Medical School}
  \city{Boston}
  \state{Massachusetts}
  \country{USA}
}
\email{kouchi@bwh.harvard.edu}

\author{Smit Desai}
\authornotemark[2]
\authornote{Corresponding author}
\orcid{0000-0001-6983-8838}
\affiliation{
  \institution{Northeastern University}
  \city{Boston}
  \state{Massachusetts}
  \country{USA}
}
\email{sm.desai@northeastern.edu}

\renewcommand{\shortauthors}{Rahman et al.}

\begin{abstract}

Serious illness conversations (SICs) align care with patients' values, goals, and preferences, yet they rarely occur in emergency departments (EDs), where time constraints and emotional burden often leave clinicians making high-stakes decisions without documented insight into what matters most to patients. We present a case study of \emph{ED GOAL-AI}, a voice-based conversational agent for brief, structured values discussions with older adults in the ED, evaluated with 55 patients for feasibility and acceptability. Most participants completed the conversation and reported the interaction as acceptable and feasible, with ratings of feeling heard and understood comparable to clinicians. However, we also observed critical failure modes, including boundary violations such as hallucinated diagnostic statements, highlighting ethical and emotional risks. This work points to early promise for AI-mediated SICs while underscoring the need for careful boundary setting and participatory design before broader deployment.

\end{abstract}

\ccsdesc[500]{Human-centered computing~Sound-based input / output}

\keywords{Large Language Models, Conversational Agents, Voice User Interfaces, Serious Illness Conversations}
\begin{teaserfigure}
  \includegraphics[height=180pt, width=\textwidth]{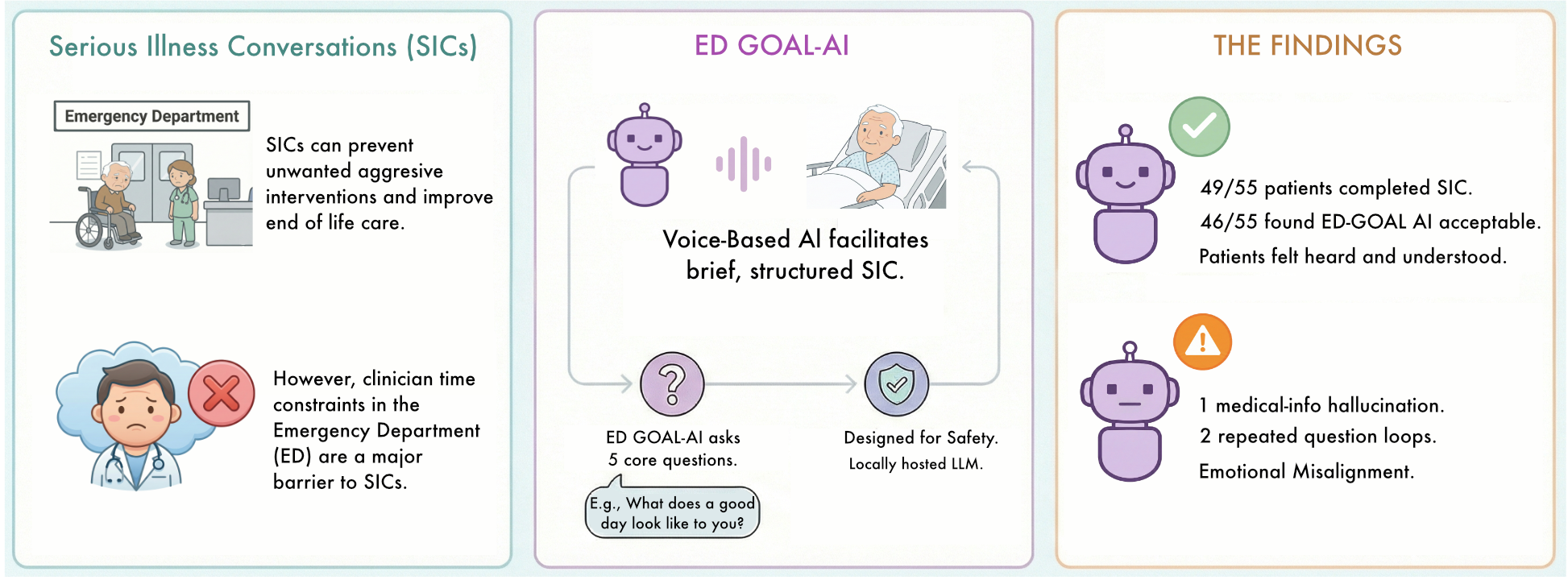}
  \caption{\textit{ED GOAL-AI} for emergency-department serious illness conversations (SICs). SICs can reduce unwanted aggressive interventions and improve end-of-life care, but clinician time constraints in the ED limit scalability (left). \textit{ED GOAL-AI} (middle) is a voice-based, locally hosted, fine-tuned LLM agent that facilitates brief, structured SICs by guiding patients through five core values questions. In a case study with patients (right), 49/55 completed the SIC and 46/55 found \textit{ED GOAL-AI} acceptable, with patients reporting feeling heard and understood; observed failure modes included one medical-information hallucination, two repeated-question loops, and emotional misalignment.}
  \Description{Three-panel infographic summarizing ED GOAL-AI for serious illness conversations. Left panel, titled “Serious Illness Conversations (SICs),” shows an emergency department scene with an older patient in a wheelchair and a clinician; accompanying text states SICs can reduce unwanted aggressive interventions and improve end-of-life care, but an illustrated clinician with a red X indicates ED time constraints are a major barrier. Middle panel, titled “ED GOAL-AI,” shows a purple robot icon communicating with a hospitalized patient via an audio waveform and arrows; text states a voice-based AI facilitates brief, structured SICs, asks five core values questions (example bubble: “What does a good day look like to you?”), and is designed for safety with a locally hosted LLM. Right panel, titled “The Findings,” lists outcomes next to icons: a green check with a robot indicates 49/55 patients completed the SIC, 46/55 found ED GOAL-AI acceptable, and patients felt heard and understood; a warning icon with a robot indicates one medical-information hallucination, two repeated-question loops, and emotional misalignment.}
  \label{fig:teaser}
\end{teaserfigure}

\maketitle

\section{Introduction}

An older adult arrives in the emergency department (ED) with acute respiratory distress. They are struggling to breathe, frightened, and meeting the ED clinician for the first time. Within minutes, the clinician must decide whether to escalate care with invasive interventions (e.g., intubation) that may prolong life at the cost of reduced quality of life, or to prioritize comfort-focused care. What is often missing in this moment is not clinical information, but documentation of prior serious illness conversations (SICs), where patients’ goals, values, and preferences for care have been discussed and recorded. In the absence of this information, clinicians are left to make high-stakes decisions without a clear understanding of what matters most to the patient.

SICs are structured discussions intended to address this gap by helping patients articulate their values, goals, and care preferences, particularly in the context of serious or life-limiting illness \cite{bernacki2014communication,houben2014efficacy,silveira2010advance}. When conducted effectively, SICs support care that aligns with patient priorities, reducing unwanted aggressive interventions and improving quality of life \cite{klingler2016does, wright2008associations}. Despite their importance, SICs remain uncommon and often occur late. Only 37\% of seriously ill older adults report having engaged in an SIC with a physician, and when these conversations do occur, they take place an average of just 33 days before death \cite{mack_associations_2012,wright2008associations}. This gap contributes to care that is misaligned with patient preferences, higher rates of in-hospital death, and increased healthcare expenditures \cite{zhang_health_2009,mack_associations_2012}.

The ED represents a critical opportunity to address this gap, but also one of the most challenging settings in which to do so. Nearly 75\% of older adults visit the ED in their final six months of life \cite{smith2012half}, often at moments when treatment trajectories are being reconsidered and goals of care become urgently relevant. Prior work has demonstrated that structured interventions such as ED GOAL can substantially increase documentation of patients’ values and goals in the ED \cite{ouchi_serious_2025}. At the same time, ED clinicians face severe time constraints and high emotional demands, making it difficult to consistently engage in extended, values-focused conversations alongside acute medical care. These constraints motivate interest in whether alternative forms of support might help facilitate SICs in emergency settings. Recent research has explored the use of large language models (LLMs) based conversational agents (CAs) for patient communication, including advance care planning and values exploration \cite{milne-ives_effectiveness_2020,huynh_applications_2026,hsu_precare_2025}. However, deploying conversational agents in moments where patients reflect on serious illness, uncertainty, and future care raises fundamental questions about feasibility, acceptability, and the appropriate role of AI in healthcare’s most sensitive interactions.

In this paper, we present a case study of \emph{ED GOAL-AI}, a voice-based conversational agent we developed to facilitate brief, structured discussions about values and goals with older adult patients in the ED of a hospital. We deployed \emph{ED GOAL-AI} with 55 older adult patients, conducting sessions at the bedside within existing ED workflows, to examine the feasibility and acceptability of AI-mediated SICs in this acute care setting. Overall, patients were able to engage with the system and generally reported these conversations as acceptable and respectful. At the same time, we observed critical limitations and failure modes, including hallucinated medical information and instances of insufficient emotional responsiveness. In sum, this work highlights both the early promise and unresolved challenges of AI-mediated SICs, and underscores the need for further investigation to establish appropriate ethical and emotional boundaries. 

\section{Related Work}

\subsection{Serious Illness Conversations in the Emergency Department}

SICs support patients in articulating their values, goals, and care preferences, enabling clinicians to align treatment decisions with what matters most to patients \cite{bernacki2014communication,houben2014efficacy,silveira2010advance}. When conducted effectively, SICs are associated with reduced unwanted aggressive interventions and improved end-of-life care outcomes \cite{klingler2016does, wright2008associations}. Rather than focusing solely on discrete medical decisions, SICs foreground broader priorities such as quality of life, acceptable trade-offs, and patients’ goals under uncertainty.

The ED represents a critical yet underutilized setting for SICs. As a primary entry point into the hospital system, the ED is often where major treatment trajectories are initiated for older adults with serious illness \cite{ouchi2019goals,prachanukool2022emergency}. Prior work has shown that timely SICs in the ED can reduce aggressive interventions and improve care alignment for patients facing life-threatening conditions \cite{ouchi2017preparing}. Despite this opportunity, SICs rarely occur in emergency settings \cite{ouchi2019goals,murray2025goc}. Clinicians face substantial barriers, including severe time constraints, averaging approximately 16 minutes of direct patient contact \cite{Reznek_Mangolds_Kotkowski_Samadian_Joseph_Larkin_2023}, and concerns about emotional burden during high-pressure encounters \cite{shilling2024let}, particularly when engaging older adults, who represent a large proportion of ED patients. Established frameworks such as the Serious Illness Conversation Guide \cite{bernacki2015development} often require more than 20 minutes to complete \cite{mandel2023pilot}, and training programs like the Serious Illness Care Program do not fully alleviate the emotional labor involved \cite{paladino2023improving}. Together, these constraints help explain why SICs remain difficult to deliver consistently in emergency care.

\subsection{Conversational Agents for Supporting Serious Illness Conversations}

In parallel, HCI and health informatics research has explored conversational agents (CAs) to support patient communication across healthcare contexts, including mental health, chronic disease management, and medication adherence \cite{lucas_reporting_2017,fitzpatrick_delivering_2017,easton_virtual_2019,piau_smartphone_2019}. Advances in LLMs have enabled more flexible dialogue and increased self-disclosure, motivating their use in higher-stakes workflows such as telehealth data collection, pre-visit intake, value exploration, and shared decision-making \cite{pas_user_2020,lee_i_2020,yang2024talk2care,li2024beyond,hsu_precare_2025,hao2024advancing}. However, LLM-based CAs remain prone to conversational failures, including inappropriate follow-ups, repetitiveness, and hallucinated information \cite{mahmood_user_2025,huang_survey_2025,lee_benefits_2023}. These risks are particularly consequential in SICs, where conversational missteps may undermine trust, emotional safety, or patient autonomy. Reflecting these concerns, prior AI work related to SICs has largely focused on patient identification or documentation \cite{avati2018improving,manz2020effect,lindvall2022natural,lee2021identifying} rather than supporting conversation itself. Recent work has begun considering whether CAs might play limited, carefully bounded roles in facilitating values-focused discussions, particularly when designed with explicit scope, constrained functionality, and clear ethical boundaries \cite{chua2022enhancing,zhao_designing_2025}. Rather than substituting for clinicians, this work frames CAs as supports for specific communication components difficult to deliver consistently in time-pressured settings. Voice-based interfaces show particular promise in hospital environments \cite{liu_scaffolded_2025}, supporting hands-free interaction while reducing barriers related to text input, screen navigation, and fine-motor demands for older adults \cite{kim_exploring_2021,bickmore_its_2005,pradhan_accessibility_2018}, who represent a large proportion of ED patients.

Building on the clinician-led ED GOAL intervention \cite{ouchi_serious_2025} and prior design work examining SIC workflow and structure in emergency care \cite{Reznek_Mangolds_Kotkowski_Samadian_Joseph_Larkin_2023,zhao_designing_2025}, this paper presents \emph{ED GOAL-AI}, a voice-based conversational agent designed to facilitate brief, structured values conversations with older adult ED patients. Rather than assuming AI-mediated SICs are appropriate, our aim is to explore whether this approach is feasible and acceptable from patients' perspectives, and how patients' experiences of feeling heard compare to clinician interactions in this context.

\section{ED GOAL-AI}
\textbf{Technical Architecture.} \emph{ED GOAL-AI} was locally hosted using a fine-tuned LLM. The fine-tuned LLM has been publicly released\footnote{https://huggingface.co/hasiburrahman/sic-qwen}. Local hosting was essential because patient conversations about end-of-life values constitute sensitive health information. While prompting can steer LLMs, in clinical settings, it is "not sufficient for all tasks" \cite{savage_fine-tuning_2025}. We fine-tuned Qwen2.5 7B\footnote{https://qwenlm.github.io/blog/qwen2.5/} using QLoRA, loading the base model in 4-bit precision and training rank-16 LoRA adapters while keeping base weights frozen. We trained on 500 synthetic conversations generated from real-world SIC transcripts and validated by clinical researchers, enabling the model to specialize in SIC dialog structure (e.g., appropriate question sequencing, acknowledgment, clarification, handling varied patient response types, and responding empathetically to emotional cues when patients express distress or share difficult information). This latter capability reflects a core palliative care communication skill, allowing the system to generate brief, supportive, and validating responses before continuing the structured conversation.

The system pipeline comprised: (1) voice activity detection, (2) speech-to-text using Whisper\footnote{https://huggingface.co/openai/whisper-medium}, (3) the locally hosted fine-tuned LLM processing transcribed text with conversation history, (4) local database storage, and (5) text-to-speech using Kokoro\footnote{https://huggingface.co/hexgrad/Kokoro-82M}. \emph{ED GOAL-AI} was voice-only without an on-screen animated persona. We tuned speech output for older adults by reducing speaking rate 10\% and increasing pause tolerance.

\textbf{Conversation Framework.} 
We adapted five core questions from the seven-item SIC guide \cite{bernacki2015development} and prior interventions, including ED GOAL \cite{ouchi_serious_2025, ouchi_empower_2019}, selecting five core questions completable within the target timeframe: (1) ``What does a good day look like for you?''; (2) ``What is your understanding of your medical problems?''; (3) ``Thinking about medical care in the future, what matters most to you?''; (4) ``Have you discussed with the people closest to you what care you would accept if you got sicker?''; and (5) ``If time becomes short, what would be most important to you then?'' \emph{ED GOAL-AI} follows a sequential structure with three stages per question: posing the question conversationally, acknowledging substantive responses before transitioning, and asking up to two clarification prompts for ambiguous or brief responses. The system prompt encodes critical constraints: role definition (compassionate conversation facilitator), safety boundaries (prohibiting prognostic, diagnostic, or treatment statements), and redirection of medical questions to clinicians with acknowledgment before transitions. Prompts are in supplementary files.

\section{Study Design}
To assess the acceptability and feasibility of \emph{ED GOAL-AI}, we conducted a single-arm study approved by the IRB at an academic tertiary-care hospital. We recruited 55 older adults from the ED in a major North American hospital system who met inclusion criteria: (1) aged $\geq$50 years with serious illness (heart failure, chronic obstructive pulmonary disease, chronic kidney disease, or metastatic cancer); (2) aged $\geq$50 with milder forms of these conditions plus hospitalization within 12 months; or (3) aged $\geq$70 years. Participants required English proficiency and the capacity to consent. Research staff screened ED patient lists to identify eligible patients and confirmed eligibility with treating clinicians. Participants had a mean age of 74.3 $\pm$
 6.6 years; 52.7\% were male, and 89.1\% were White. Twenty participants qualified through serious illness pathways (metastatic cancer 16.4\%, heart failure 14.5\%, CKD 14.6\%, COPD 3.6\%), while 35 qualified by age. Our sample size and eligibility criteria were consistent with prior work exploring SIC feasibility in the ED \cite{ouchi_empower_2019, pajka_feasibility_2021}.

\begin{figure}[ht]
\centering
\includegraphics[scale=0.14]{imgs/ED-Goal.png}
\caption{A patient uses \textit{ED GOAL-AI} on a tablet in the ED while research staff remains present for technical support without prompting or directing the discussion. Photo shared with patient's permission; identifying details have been redacted.}
\Description{Photograph in an emergency department exam room. A patient lies in a hospital bed on the right, wearing a patterned gown, facing a tablet. On the left, a masked research staff member sits nearby and holds a tablet upright toward the patient, as if facilitating the use of the system while not engaging in the conversation. The patient’s and staff member’s faces and a name badge are covered by black rectangles. Medical equipment and cables, a computer keyboard, and an IV pole are visible in the background. A banner at the top reads that the patient gave permission to share the photo.}
\label{fig:procedure}
\end{figure}

\textbf{\textit{Procedure.}} With the treating clinician’s permission, research staff approached eligible patients, explained the study, and obtained verbal consent. Enrolled participants completed a single conversation with ED GOAL-AI lasting approximately 5 minutes, a duration informed by the efficacy of brief negotiated interviews in ED \cite{donofrio_development_2005, bernstein_project_1997, donofrio_brief_2012} and the need to integrate without disrupting clinical workflows in a time-sensitive context \cite{rubin_refinement_2022, ouchi_empower_2019}. Interactions occurred via tablet device, with the patient speaking to the CA and hearing responses through the device speaker. This was consistent with the prior ED GOAL intervention, in which nurses communicated with patients via a tablet held by research staff \cite{ouchi_serious_2025}. Research staff remained present to provide technical support and intervene if necessary (Figure \ref{fig:procedure}), but did not prompt participants or guide conversation content. After the interaction, participants completed a post-interaction survey. If the participant expressed anxiety, distress, or a critical failure mode occurred, they could discontinue the study, and an SIC-trained physician would address their emotional needs.

\textbf{\textit{Measures.}} The primary outcome was acceptability, assessed using four 5-point items measuring acceptability, respectfulness, clarity, and ease of interaction, consistent with prior ED-based SIC feasibility studies \cite{ouchi_empower_2019, pajka_feasibility_2021}. Secondary outcomes included feasibility (completion rate, conversation duration, interruptions, and non-responses) and the NQF-validated five-point Feeling Heard and Understood survey \cite{gramling_feeling_2016}, which asked participants to rate how heard and understood they felt by their physicians and by \emph{ED GOAL-AI} regarding what mattered in their care.

\section{Findings}

\begin{figure*}[ht]
\centering
\includegraphics[scale=1.3]{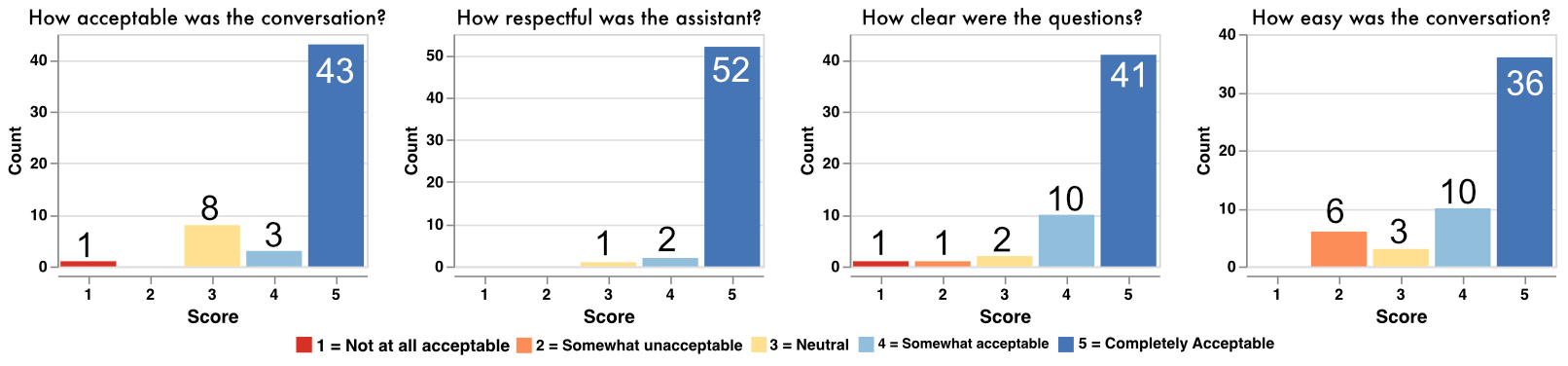}
\caption{Acceptability ratings for \textit{ED GOAL-AI} across four dimensions (1--5 Likert; higher is more acceptable): acceptability, respectfulness, question clarity, and ease of conversation. Most participants rated the \textit{ED GOAL-AI} as completely acceptable.}
\Description{Four small-multiple bar charts showing counts of 1--5 Likert ratings (x-axis: Score; y-axis: Count) for ED GOAL-AI. Panel 1, “How acceptable was the conversation?”: 43 ratings of 5, 3 ratings of 4, 8 ratings of 3, and 1 rating of 1. Panel 2, “How respectful was the assistant?”: 52 ratings of 5, 2 ratings of 4, and 1 rating of 3. Panel 3, “How clear were the questions?”: 41 ratings of 5, 10 ratings of 4, 2 ratings of 3, 1 rating of 2, and 1 rating of 1. Panel 4, “How easy was the conversation?”: 36 ratings of 5, 10 ratings of 4, 3 ratings of 3, and 6 ratings of 2. Across all panels, the distribution is concentrated at score 5, indicating high perceived acceptability.}
\label{fig:acceptability}
\end{figure*}

\textbf{\textit{Acceptability.}} Participants rated \emph{ED GOAL-AI} positively across all four acceptability dimensions (Figure \ref{fig:acceptability}). Respectfulness received the highest ratings, with 52 participants (94.5\%) rating the system as ``completely respectful''. Acceptability and clarity showed similar patterns, with over 90\% of participants providing positive ratings (``somewhat'' or ``completely'' acceptable). Ease of conversation exhibited the most variation, with 36 participants (65.5\%) rating it ``completely acceptable '', potentially reflecting the inherent difficulty of discussing end-of-life topics rather than interface usability. 

\textbf{\textit{Feasibility.}} Forty-nine participants (89.1\%) completed conversations successfully. Mean duration was $M = 4.5$ minutes ($SD = 2.3$), aligning with our five-minute target, with $Mdn = 6$ questions asked ($IQR = 5$--$9$). Older adults were generally able to engage with the voice interface, with only two interruptions and three non-response events observed. Six conversations (10.9\%) were not completed: four due to early technical issues (one hallucination, two question repetitions, and one audio failure), one at a participant’s request, and one due to clinical interruption. The technical issues occurred during the first 23 interactions and were traced to improper management of the LLM’s key–value cache, which maintains short-term conversational context during multi-turn dialogue. After revising cache management, no further technical failures were observed.

\textbf{\textit{Feeling Heard and Understood.}} No participants reported feeling ``not at all'' heard by \emph{ED GOAL-AI}. While primary doctors received slightly higher ratings ($M = 4.49$, $SD = 0.81$) than \emph{ED GOAL-AI} ($M = 4.25$, $SD = 0.95$), this difference was not significant ($p = .129$).

\textbf{\textit{Observational Notes. }}Research staff noted instances where \emph{ED GOAL-AI}'s responses were misaligned with patients' emotional states. In several early encounters, when seriously ill patients became tearful, \emph{ED GOAL-AI} responded with phrases like ``I completely understand'' rather than explicitly acknowledging the emotion. In clinical practice, clinicians are taught to avoid such statements and instead use language like ``I can only imagine'' to validate emotion without implying shared lived experience \cite{childers_beyond_2023, back_teaching_2003}. These interactions underscored that small phrasing choices can meaningfully shape perceived empathy and alignment in SIC, even when the system otherwise supports portions of the conversation.

\section{Discussion}

Our findings indicate early promise for voice-based conversational agents to support values-focused reflection in emergency care. In a setting where time pressure and emotional burden often prevent clinicians from initiating SICs, a brief, structured AI-facilitated interaction eliciting meaningful discussion about goals and values is notable. This aligns with prior HCI work arguing that AI systems add value when positioned as assistive supports rather than replacements for human judgment or care \cite{zhang2024rethinking, zhao_designing_2025}. However, our study surfaced important failure modes that complicate any simple notion of success. We observed a hallucination event where \emph{ED GOAL-AI} generated diagnostic statements despite being explicitly designed not to provide diagnosis, prognosis, or treatment recommendations. These incidents highlight how conversational AI failures in serious illness contexts are best understood as boundary violations rather than general performance errors. Prior HCI work has cautioned that conversational agents in healthcare may inadvertently project authority or understanding beyond their actual competence, creating an "illusion" of empathy or expertise \cite{cuadra2024illusion,seitz2024artificial,kong2024envisioning}. We extend these concerns by showing that even rare boundary crossings carry outsized ethical and emotional consequences in acute care settings.

Our case study highlights both the potential and unresolved challenges of AI-mediated SICs. While overall feasibility and acceptability suggest AI may play a limited role in facilitating values discussions in the ED, our observations underscore that feasibility alone is insufficient. Responsible progress requires explicit attention to ethical and emotional boundaries, robust constraints on system behavior, and participatory design with patients and clinicians to determine what AI should and should not do in moments of acute vulnerability \cite{Bainbridge_1983,Weiser1996THECA}. We position AI-mediated SICs as a high-stakes human-centered design problem where early promise must be evaluated alongside the risks of rare but consequential failures, and where safeguarding patient autonomy and trust is as critical as technical advancement.

\section{Limitations and Future Work}
This study was conducted in a well-resourced hospital ED, which may limit generalizability to settings with different resources, workflows, or patient populations. Future work should examine how AI-mediated SICs translate to other hospital contexts. Future iterations of \emph{ED GOAL-AI} should incorporate greater exposure to real-world clinical transcripts and training aligned with established SIC emotion-response frameworks, prioritizing emotional attunement and conversational appropriateness beyond factual accuracy. Our deployment faced on-device computational constraints that limited fully natural conversational dynamics, suggesting the need for interaction protocols and safeguards aligned with current system capabilities. Finally, participatory co-design with patients, patient advisors, and clinicians will be essential to establish ethical and emotional boundaries and determine where such systems can appropriately fit within clinical workflows.


\begin{acks}
We thank the members of the Conversational Human-AI Interactions (CHAI) Lab for their support, feedback, and discussions throughout this work. We also thank Dr. Shinichi Goto for his valuable contributions. We are grateful to the clinical team and staff at Brigham and Women's Hospital for their support. Above all, we are deeply grateful to the patients who participated in the study, whose generosity and trust made this case study possible. Generative AI was used to generate components of Figure \ref{fig:teaser}.
\end{acks}

\bibliographystyle{ACM-Reference-Format}
\bibliography{ref}


\end{document}